%% file: MACS0647.tex
\documentclass[twocolumn]{aastex63}

\shorttitle{Lensed star candidates in MACS0647}
\shortauthors{A. K. Meena et al.}

\usepackage{txfonts}
\usepackage[T1]{fontenc}
\usepackage{longtable}

\begin{document}

\title{Two lensed star candidates at $z\simeq4.8$ behind the galaxy cluster MACS J0647.7+7015}


\correspondingauthor{Ashish Kumar Meena}
\email{ashishmeena766@gmail.com}

\author[0000-0002-7876-4321]{Ashish Kumar Meena}
\affiliation{Physics Department, Ben-Gurion University of the Negev, P.O. Box 653, Be'er-Sheva 84105, Israel}

\author[0000-0002-0350-4488]{Adi Zitrin}
\affiliation{Physics Department, Ben-Gurion University of the Negev, P.O. Box 653, Be'er-Sheva 84105, Israel}

\author[0000-0002-6090-2853]{Yolanda Jim\'enez-Teja}
\affiliation{Instituto de Astrof\'isica de Andaluc\'ia, Glorieta de la Astronom\'ia s/n, 18008 Granada, Spain}
\affiliation{Observatório Nacional - MCTI (ON), Rua Gal. José Cristino 77, São Cristóvão, 20921-400, Rio de Janeiro, Brazil}

\author[0000-0003-1096-2636]{Erik Zackrisson}
\affiliation{Observational Astrophysics, Department of Physics and Astronomy, Uppsala University, Box 516, SE-751 20 Uppsala, Sweden}

\author[0000-0003-1060-0723]{Wenlei Chen}
\affiliation{School of Physics and Astronomy, University of Minnesota, 116 Church Street SE, Minneapolis, MN 55455, USA}

\author[0000-0001-7410-7669]{Dan Coe}
\affiliation{Space Telescope Science Institute (STScI), 3700 San Martin Drive, Baltimore, MD 21218, USA}
\affiliation{Association of Universities for Research in Astronomy (AURA) for the European Space Agency (ESA), STScI, Baltimore, MD, USA}
\affiliation{Center for Astrophysical Sciences, Department of Physics and Astronomy, The Johns Hopkins University, 3400 N Charles St. Baltimore, MD 21218, USA}

\author[0000-0001-9065-3926]{Jose M. Diego}
\affiliation{Instituto de F\'isica de Cantabria (CSIC-UC). Avda. Los Castros s/n. 39005 Santander, Spain}  

\author[0000-0001-7399-2854]{Paola Dimauro}
\affiliation{INAF - Osservatorio Astronomico di Roma, via di Frascati 33, 00078 Monte Porzio Catone, Italy}

\author[0000-0001-6278-032X]{Lukas J. Furtak}
\affiliation{Physics Department, Ben-Gurion University of the Negev, P.O. Box 653, Be'er-Sheva 84105, Israel}

\author[0000-0003-3142-997X]{Patrick L. Kelly}
\affiliation{School of Physics and Astronomy, University of Minnesota, 116 Church Street SE, Minneapolis, MN 55455, USA}

\author[0000-0003-3484-399X]{Masamune Oguri}
\affiliation{Center for Frontier Science, Chiba University, 1-33 Yayoi-cho, Inage-ku, Chiba 263-8522, Japan}
\affiliation{Department of Physics, Graduate School of Science, Chiba University, 1-33 Yayoi-Cho, Inage-Ku, Chiba 263-8522, Japan}

\author[0000-0003-1815-0114]{Brian Welch}
\affiliation{Department of Astronomy, University of Maryland, College Park, MD 20742, USA}
\affiliation{Observational Cosmology Lab, NASA Goddard Space Flight Center, Greenbelt, MD 20771, USA}
\affiliation{Center for Research and Exploration in Space Science and Technology, NASA/GSFC, Greenbelt, MD 20771}

\author[0000-0002-5258-8761]{Abdurro'uf}
\affiliation{Center for Astrophysical Sciences, Department of Physics and Astronomy, The Johns Hopkins University, 3400 N Charles St. Baltimore, MD 21218, USA}
\affiliation{Space Telescope Science Institute (STScI), 3700 San Martin Drive, Baltimore, MD 21218, USA}

\author[0000-0002-8144-9285]{Felipe Andrade-Santos}
\affiliation{Department of Liberal Arts and Sciences, Berklee College of Music, 7 Haviland Street, Boston, MA 02215, USA}
\affiliation{Center for Astrophysics \text{\textbar} Harvard \& Smithsonian, 60 Garden Street, Cambridge, MA 02138, USA}

\author[0000-0002-0786-7307]{Angela Adamo}
\affiliation{Department of Astronomy, Oskar Klein Centre, Stockholm University, AlbaNova University
Centre, SE-106 91 Stockholm, Sweden}

\author[0000-0003-0883-2226]{Rachana Bhatawdekar}
\affiliation{European Space Agency, ESA/ESTEC, Keplerlaan 1, 2201 AZ Noordwijk, The Netherlands}

\author[0000-0001-5984-0395]{Maru\v{s}a Brada\v{c}}
\affiliation{Department of Mathematics and Physics, University of Ljubljana, Jadranska ulica 19, SI-1000 Ljubljana, Slovenia}
\affiliation{Department of Physics and Astronomy, University of California, Davis, 1 Shields Ave, Davis, CA 95616, USA}

\author[0000-0002-7908-9284]{Larry D. Bradley}
\affiliation{Space Telescope Science Institute (STScI), 3700 San Martin Drive, Baltimore, MD 21218, USA}

\author[0000-0002-8785-8979]{Tom Broadhurst}
\affiliation{Department of Theoretical Physics, University of the Basque Country UPV/EHU, Bilbao, Spain}
\affiliation{Donostia International Physics Center (DIPC), 20018 Donostia, Spain}
\affiliation{IKERBASQUE, Basque Foundation for Science, Bilbao, Spain}

\author[0000-0003-1949-7638]{Christopher J. Conselice}
\affiliation{Jodrell Bank Centre for Astrophysics, University of Manchester, Oxford Road, Manchester UK}

\author[0000-0001-8460-1564]{Pratika Dayal}
\affiliation{Kapteyn Astronomical Institute, University of Groningen, P.O. Box 800, 9700 AV Groningen, The Netherlands}

\author[0000-0002-2808-0853]{Megan Donahue}
\affiliation{Michigan State University, Physics \& Astronomy Department, East Lansing, MI, USA}

\author[0000-0003-1625-8009]{Brenda L.~Frye}
\affiliation{Department of Astronomy/Steward Observatory, University of Arizona, 933 N. Cherry Avenue, Tucson, AZ 85721, USA}

\author[0000-0001-7201-5066]{Seiji Fujimoto}\altaffiliation{Hubble Fellow}
\affiliation{Department of Astronomy, The University of Texas at Austin, Austin, TX 78712, USA}

\author[0000-0003-4512-8705]{Tiger Yu-Yang Hsiao}
\affiliation{Center for Astrophysical Sciences, Department of Physics and Astronomy, The Johns Hopkins University, 3400 N Charles St. Baltimore, MD 21218, USA}

\author[0000-0002-5588-9156]{Vasily Kokorev}
\affiliation{Kapteyn Astronomical Institute, University of Groningen, PO Box 800, 9700 AV Groningen, The Netherlands}

\author[0000-0003-3266-2001]{Guillaume Mahler}
\affiliation{Centre for Extragalactic Astronomy, Durham University, South Road, Durham DH1 3LE, UK}
\affiliation{Institute for Computational Cosmology, Durham University, South Road, Durham DH1 3LE, UK}

\author[0000-0002-5057-135X]{Eros Vanzella}
\affiliation{INAF -- OAS, Osservatorio di Astrofisica e Scienza dello Spazio di Bologna, via Gobetti 93/3, I-40129 Bologna, Italy}

\author[0000-0001-8156-6281]{Rogier A. Windhorst} 
\affiliation{School of Earth and Space Exploration, Arizona State University,
Tempe, AZ 85287-1404, USA}

\begin{abstract}
We report the discovery of two extremely magnified lensed star candidates behind the galaxy cluster MACS~J0647.7+7015 using recent multi-band \textit{James Webb Space Telescope} (\emph{JWST}) NIRCam observations. The star candidates are seen in a previously known, $z_{phot}\simeq4.8$ dropout giant arc that straddles the critical curve. The candidates lie near the expected critical curve position, but lack clear counter images on the other side of it, suggesting these are possibly stars undergoing caustic crossings. We present revised lensing models for the cluster, including multiply imaged galaxies newly identified in the JWST data, and use them to estimate a background macro-magnification of at least $\gtrsim90$ and $\gtrsim50$ at the positions of the two candidates, respectively. With these values, we expect effective, caustic-crossing magnifications of~$\sim[10^3-10^5]$ for the two star candidates. The Spectral Energy Distributions (SEDs) of the two candidates match well spectra of B-type stars with best-fit surface temperatures of $\sim10,000$~K, and $\sim12,000$~K, respectively, and we show that such stars with masses $\gtrsim20$ M$_{\odot}$ and $\gtrsim50$ M$_{\odot}$, respectively, can become sufficiently magnified to be observable. We briefly discuss other alternative explanations and conclude these objects are likely lensed stars, but also acknowledge that the less magnified candidate may alternatively reside in a star cluster. These star candidates constitute the second highest-redshift examples to date after Earendel at $z_{phot}\simeq6.2$, establishing further the potential of studying extremely magnified stars to high redshifts with the \emph{JWST}. Planned future observations, including with NIRSpec will enable a more detailed view of these candidates in the near future.  
\end{abstract}


\section{Introduction}
\label{sec:intro}

The serendipitous discovery by \citet{Kelly2018Icarus} several years ago of the first highly magnified star in \emph{Hubble Space Telescope~(HST)} imaging in the MACS~J1149.5+2223 galaxy cluster~\citep[$z=0.544$;][]{2007ApJ...661L..33E}, has opened a new window to observe stars at cosmological distances~\citep[e.g.,][]{1991ApJ...379...94M}. The star~\citep[named `Icarus';][]{Kelly2018Icarus} was detected in a strongly lensed spiral galaxy~($z=1.49$) and was found to have an estimated magnification factor of~$\sim2000$. Several other lensed stars were since detected in \emph{HST} imaging of various galaxy clusters~\citep{Rodney2018Spock, Chen2019Warhol, Kaurov2019Warhol, Welch2022EarendelHST, Diego2022Godzilla, 2022arXiv221101402M}, with rapidly increasing numbers \citep{2022arXiv221102670K}. Thanks to the larger photon collecting area compared to \emph{HST}, and its sensitivity to infrared light, the \emph{JWST}~\citep{2006SSRv..123..485G} significantly enhances our ability to detect such lensed stars, especially at higher redshifts \citep[][]{Windhorst2018, 2022MNRAS.514.2545M}. So far, nearly all galaxy clusters observed by \emph{JWST} revealed lensed stars~\citep[][and several more are forthcoming]{Pascale2022SMACS0723, Chen2022, Diego2022Quyllur, Welch2022EarendelJWST}, showcasing the promising rate of such detections.

Observing highly magnified stars at cosmological distances is typically a result of a combined effect of strong- and micro-lensing~\citep[e.g.,][]{1991ApJ...379...94M, Oguri2018}. The presence of point-like masses (such as stellar-mass objects) in the lens leads to the formation of micro critical curves and micro-caustics in the lens and source planes respectively. The area covered by each of these micro critical curves depends on the perturber's mass and on the macro-magnification (the higher the mass and macro-magnification, the larger the area) and at sufficiently high macro magnifications the micro-critical curves merge with each other to form a corrugated network~\citep[e.g.,][]{2017ApJ...850...49V, Diego2018, Diego2019}. Whenever a compact source, such as e.g.~a star in a strongly lensed galaxy, crosses a micro-caustic, it gets highly magnified -- leading to a peak in its light curve, which might make it observable for a brief period of time as a transient source. The height, width and frequency of these peaks depend on various factors such as the relative velocity and radius of the source, the surface density of microlenses, and their mass function~\citep[e.g.,][]{2017ApJ...850...49V}. If the underlying magnification is sufficiently high, and the corrugated micro-caustic network is dense enough, highly magnified stars may be seen as persistent sources with only moderate fluctuations~\citep[e.g.,][]{Welch2022EarendelHST, Welch2022EarendelJWST}.

The substantial magnification provided by galaxy cluster lenses has also continuously led to the detection of background high-redshift galaxy candidates in observations with the \emph{HST}~\citep[e.g.,][]{2013ApJ...762...32C, 2017ApJ...835...44C, 2018ApJ...864L..22S, 2019MNRAS.486.3805B} and \emph{JWST}~\citep[e.g.,][]{2022arXiv220711217A, 2022arXiv220712338A, 2022ApJ...938L..15C, furtak22b, 2022arXiv221014123H, 2022arXiv221109839V, 2022arXiv221015699W, 2022arXiv220711558Y}. One of these candidates is MACS0647-JD~\citep{2013ApJ...762...32C}, a record-breaking, triply imaged galaxy at $z\simeq10.7$, which was first detected in \emph{HST} imaging of the MACSJ~0647.7+7015~(hereafter MACS0647; $z=0.591$, \citealt{2007ApJ...661L..33E}) galaxy cluster under the \emph{Cluster Lensing and Supernova survey with Hubble~(CLASH)}\footnote{\url{https://www.stsci.edu/~postman/CLASH/}} program~\citep{2012ApJS..199...25P}. To study MACS0647-JD in more detail, the \emph{JWST} general observer (GO) program~1433 (PI: Dan Coe) targeted the cluster and obtained \emph{JWST}/NIRCam imaging in six different filters. The \emph{JWST} imaging supported the high-redshift nature at~$z_{phot}=10.6$ with very high-confidence, and revealed that MACS0647-JD is actually made of two components with distance~$\sim400$~pc, and a possible third clump about 3 kpc away \citep{2022arXiv221014123H}. 

In this work, we present and study the properties of two extremely magnified lensed star candidates identified in these \emph{JWST}/NIRCam observations of MACS0647. The candidates are seen in a giant arc at a redshift~$z_{\rm phot}=4.79_{-0.15}^{+0.07}$ lensed by the cluster. They are identified primarily by their compactness, their position in the arc, and their proximity to the critical curve -- implying very high background magnifications; and the lack of counter images -- implying they are possibly experiencing a local temporary extreme magnification, or alternatively sit sufficiently close to the macro-caustic so that their two counter images are merged into one, single unresolved image. 

This paper is organized as follows: In section~\ref{sec:jwstdata}, we briefly describe the \emph{JWST}/NIRCam imaging of MACS0647. In Section~\ref{sec:models}, we present revised strong lens models for MACS0647, needed for the interpretation of the sources. In Section~\ref{sec:LensedStars}, we discuss the highly magnified lensed star candidates. The cosmological parameters used in this work to estimate the various parameters are:~$H_0=70\:{\rm km\:s^{-1}\:Mpc^{-1}}$, $\Omega_{m}=0.3$, and $\Omega_{\Lambda}=0.7$. With these, $1''$ corresponds to 6.37\,kpc at the cluster redshift. All magnitudes are in the AB system~\citep{1983ApJ...266..713O}.

\section{\emph{JWST} data}
\label{sec:jwstdata}

MACS0647 was observed by \emph{JWST} in September~2022 as part of a cycle 1 GO program (program ID: 1433; PI: Dan Coe). All of the corresponding data products are publicly available on the \emph{Mikulski Archive for Space Telescopes (MAST)} website at the Space Telescope Science Institute. The specific observations analyzed can be accessed via:~\dataset[10.17909/d2er-wq71]{http://dx.doi.org/10.17909/d2er-wq71}. Within this program, MACS0647 was observed in the six NIRCam filters: F115W, F150W, F200W, F277W, F356W, and F444W spanning the wavelength range from~${\sim}~1\mu{\rm m}$ to ${\sim}~5\mu{\rm m}$\footnote{\url{https://s3.amazonaws.com/grizli-v2/JwstMosaics/v4/index.html}}. In each filter, the total exposure time was 2104 seconds, reaching a $5\sigma$ limiting magnitude in the range~28 to~29~AB. We note that additional imaging in F200W and F480M, along with NIRSpec spectroscopic observations, are expected to be conducted in early 2023 on this target.

In this work we use images reduced with the \texttt{GRIZLI}~software \citep{Grizli} and photometric redshifts estimated using \texttt{EAZY}~\citep{2008ApJ...686.1503B}. In addition to the \textit{JWST} data, \emph{HST} observations in 16 ACS and WFC3 filters covering UV to the near-infrared wavelengths are also used, mainly from the \textit{CLASH} program (GO 12101; PI: Postman), including previous data from GO~9722~(PI: Ebeling), GO~10493, 10793~(PI: Gal-Yam), and GO~13317 (PI: Coe) programs, \citep[see][]{2007ApJ...661L..33E, 2012ApJS..199...25P, 2013ApJ...762...32C}. 

For more details about the data and their reduction procedure, we refer the reader to \citet{2022arXiv221014123H}.

\begin{figure*}
    \centering
    \includegraphics[scale=0.59]{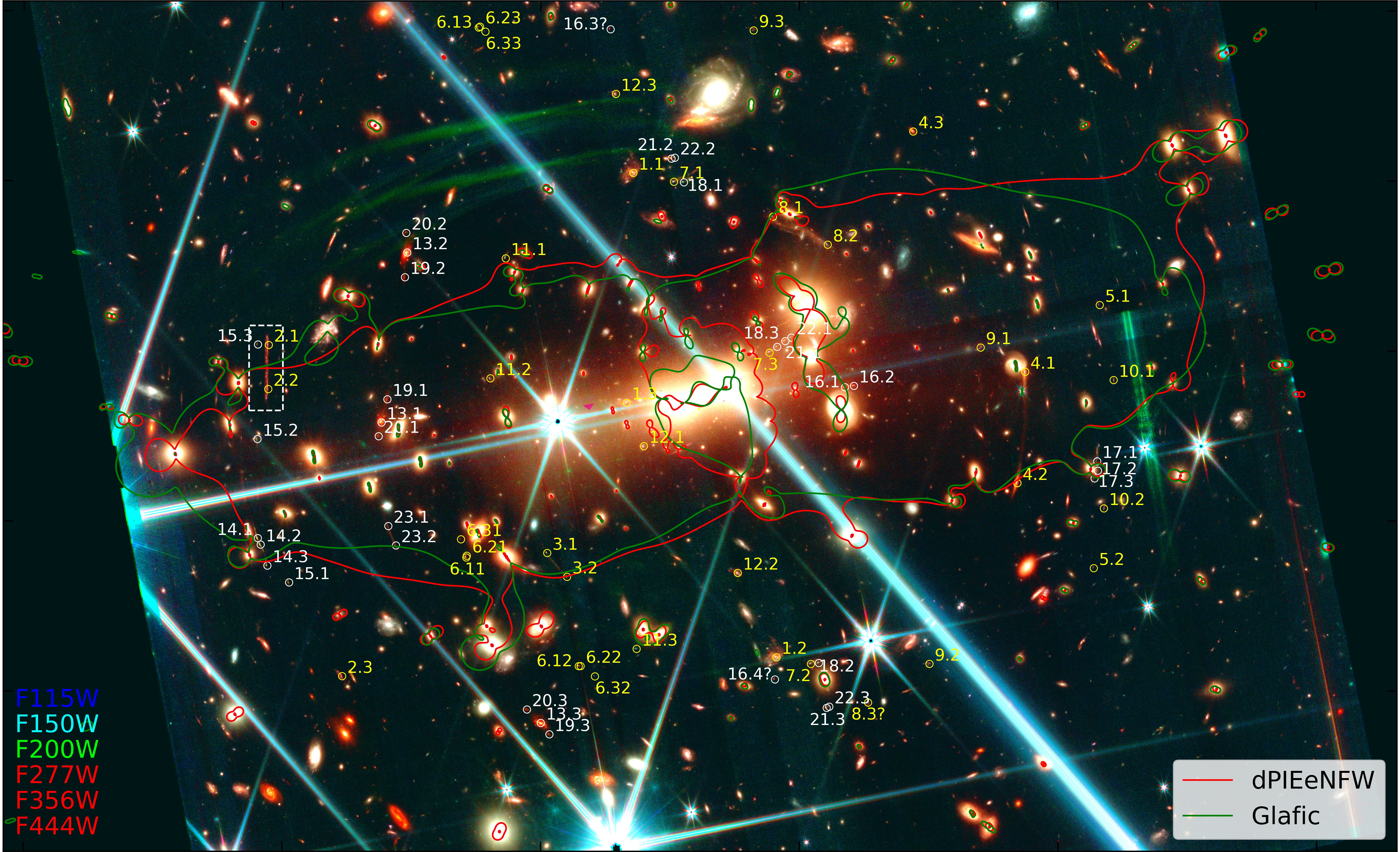}
    \caption{\emph{JWST}/NIRCam false color image (R=F277W+F356W+F444W; G=F150W+F200W; B=F115W+F150W) of MACS0647. Multiple images are indicated and numbered. Yellow circles mark lensed image systems previously identified in \emph{HST} observations, and white circles denote lensed systems newly identified with \emph{JWST}. The red and green curves represent the macro critical curves corresponding to our \texttt{dPIEeNFW} and \texttt{Glafic} lens models, for a source at redshift~$z_{s}=4.8$. The white dashed box marks the position of the `giant arc' (resulting from the merger of images 2.1 and 2.2) in which the lensed star candidates are detected. The panel size is~$2\farcm73\times1\farcm64$ across.}
    \label{fig:M0647_CC_LI}
\end{figure*}

\begin{figure*}
    \centering
    \includegraphics[scale=1.0]{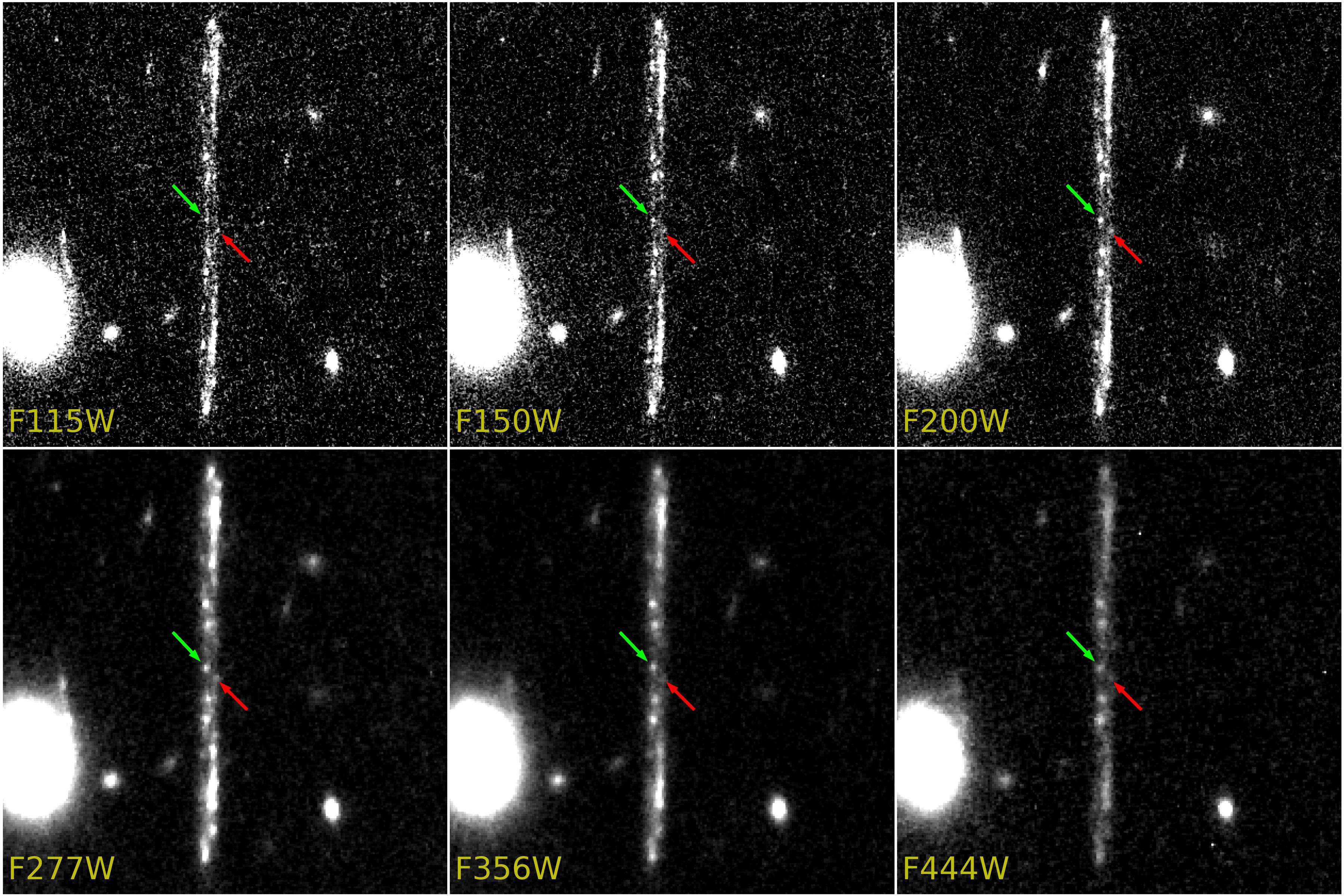}
    \caption{The two lensed star candidates in different \emph{JWST} filters. In each panel, the positions of \texttt{star-1} and \texttt{star-2} on the lensed arc are shown by green and red arrows, respectively. Each panel is~$8\farcs4\times8\farcs4$ across.}
    \label{fig:z5_stamp}
\end{figure*}

\begin{figure}
    \centering
    \includegraphics[scale=0.55]{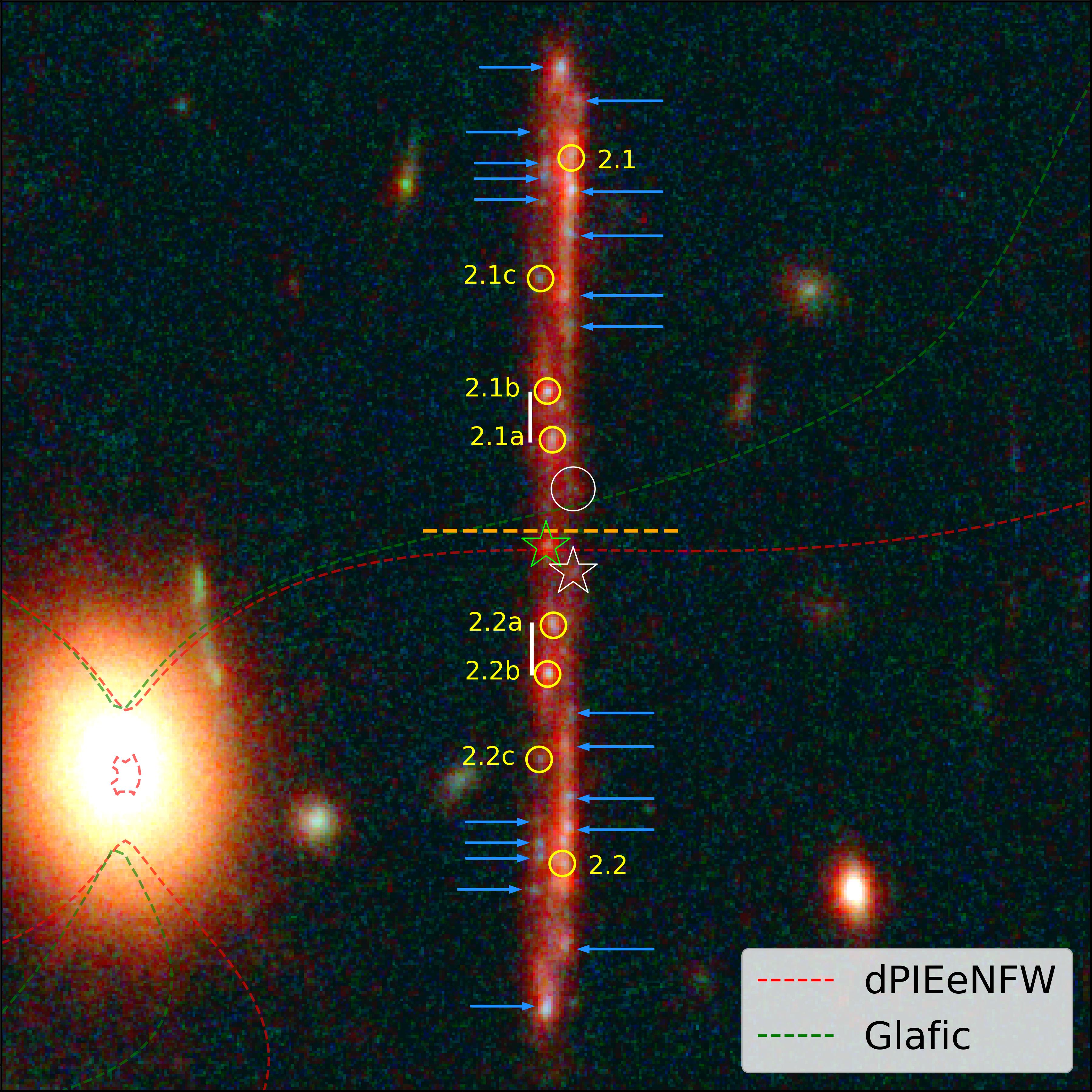}
    \caption{Highly magnified star candidates in the strongly lensed arc at~$z_s=4.8$. The dashed red and green curves represent the critical curve corresponding to the \texttt{dPIEeNFW} and \texttt{Glafic} lens mass models, respectively. The positions of \texttt{star-1} and \texttt{star-2} are marked by the green and white stars, respectively. The dashed yellow line perpendicular the arc show the position of the macro-critical curve on the arc estimated model-independently using the distances between the counter-images of two strongly lensed knots. These knots sit at the endpoints of the white lines drawn along the arc. The blue arrows show various multiply imaged clumps in the galaxy. The yellow-circled clumps are the only one used in the \texttt{dPIEeNFW} lens model reconstruction. The white circle marks the position of a fainter, possible counter image of \texttt{star-2}. The panel is~$8\farcs4\times8\farcs4$ across.}
    \label{fig:z5_color}
\end{figure}

\begin{figure}
    \centering
    \includegraphics[scale=0.55]{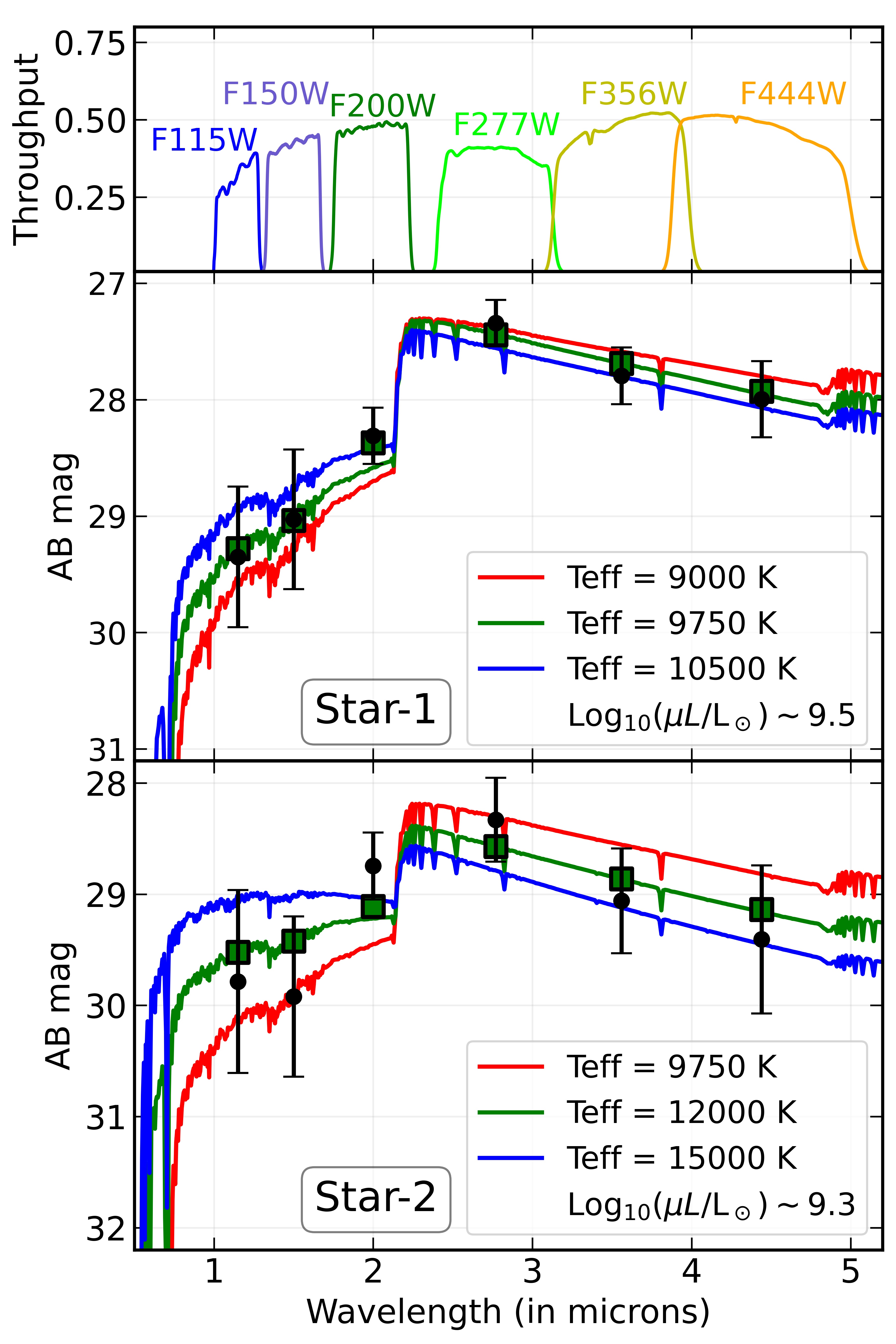}
    \caption{Measured photometry of \texttt{star-1} and \texttt{star-2} and the corresponding Spectral Energy Distribution~(SED) model fits are shown in middle and bottom panels, respectively. The black solid points with error-bars represent the observed AB magnitudes and their ~$1\sigma$ errors for the lensed star candidates, in the various \emph{JWST}/NIRCam filters. The red, green, and blue curves for each candidate represent three acceptable single-star SED fits, with the green SED being the best-fitting one. For the green SEDs, we also plot the integrated broadband fluxes (filled squares) resulting from the model and present the best-fitting $\log(\mu L/L_\odot)$ scaling of this fit. The filter throughput curves are shown on the top panel.}
    \label{fig:star1}
\end{figure}

\section{Strong Lens Modeling of MACS0647}
\label{sec:models}

Strong lens models for MACS0647 have been constructed in the past based on \emph{HST} observations. The first preliminary lens model for MACS0647 was presented in~\citet{2011MNRAS.410.1939Z} based on the two multiple image systems found in pre-\emph{CLASH}, \emph{HST/ACS} F555W+F814W imaging, using the \texttt{Light-Trace-Mass}~\citep[\texttt{LTM;}][]{2009MNRAS.396.1985Z} method. In \citet{2013ApJ...762...32C} seven new multiple-image systems were identified in the \emph{CLASH} observations, including MACS0647-JD at $z\simeq10.6$. Based on the previous and newly identified systems~\citet{2013ApJ...762...32C} refined the previous \texttt{LTM} lens model and presented two additional lens models using~\texttt{Lenstool}~\citep[parameteric;][]{1993A&A...273..367K,2007NJPh....9..447J} and \texttt{LensPerfect}~\citep[non-parametric;][]{2008ApJ...681..814C}. \citet{2015ApJ...801...44Z} identified three additional strongly lensed system candidates, bringing the total number of strongly lensed systems know in this cluster to twelve in the pre-\emph{JWST} observations. Furthermore, a \texttt{Glafic}~\citep{2010PASJ...62.1017O} lens model for MACS0647 was discussed in~\citet{2020MNRAS.496.2591O}.

Thanks to its superior capabilities in (near-)infrared resolution and depth, \emph{JWST} brings forth a large number of new lensed sources. We have visually inspected the \emph{JWST} images and identified~11 new strongly lensed system candidates -- and following similar symmetry -- in addition to the~12 already known. The complete list of the 23 multiple image systems is given here in Table~\ref{tab:lensed}. In addition to the newly identified lensed image candidates, we also detect multiple pairs of small scale substructures within some of these lensed images. For example, system~6, according to the \emph{HST} images was an isolated triply-imaged galaxy. However, \emph{JWST} imaging reveals that it consists of two clumps with a separation of~$\sim400$~pc from each other~\citep{2022arXiv221014123H}. Similarly, in the giant lensed arc at $z\simeq4.8$~(system~2), we observe multiple pairs of strongly lensed stellar clumps in the source~(see Figure~\ref{fig:z5_color}).

We here construct revised lens models for MACS0647 using our~\texttt{dPIEeNFW}~\citep[see][]{2015ApJ...801...44Z} and~\texttt{Glafic}~\citep{2010PASJ...62.1017O} codes. These new mass models are based on the earlier known multiple-image systems, as well as the newly found ones. To our knowledge, none of the multiple image systems have a spectroscopic redshift measurement. However, the dropout nature of the $z\simeq4.8$ arc and the JD object, for example, which also span a large range of lensing distance ratios, together with tightly constrained multi-band~(16~\emph{HST} + 6~\emph{JWST}) based photometric redshifts for most systems, allow us to construct robust lens models.  

Both of the lens models used here are parametric in nature. The \texttt{dPIEeNFW} implementation which we use is a revised version of the parametric code~\citet{2015ApJ...801...44Z} used to map the \emph{CLASH} sample and the \emph{Hubble Frontier Fields}. The main improvement is that the new version is not grid-based and thus can reach higher resolutions and gives more accurate results. The new method has already been implemented on various clusters with \emph{JWST} data~\citep[e.g.,][]{Pascale2022SMACS0723, 2022arXiv221015639R, 2022arXiv221014123H, 2022arXiv221015699W}\footnote{referred to as ``\texttt{Zitrin-Analytic}'' in some of these studies}. Here we use 175 cluster members chosen by the red sequence of the cluster and parametrized as double pseudo isothermal elliptical mass-density profiles, and two cluster-scale DM halos each parameterized as an elliptical NFW~\citep{1996ApJ...462..563N}.  The centers of these are optimized in the minimization procedure, around the potions of the central BCGs. Minimization is done in the source plane~\citep{2010GReGr..42.2151K}, via a several-dozen thousand step Markov Chain Monte Carlo (MCMC), from which the uncertainties are derived as well. The \texttt{GLAFIC} modeling code~\citep{2010PASJ...62.1017O} has been successfully applied to a large sample of clusters before, and was proven very robust when compared to numerically simulated clusters, or for time-delay predictions~\citep[e.g.,][]{2016ApJ...819L...8K, 2017MNRAS.472.3177M}. Pseudo-Jaffe profiles are used to describe the galaxies whereas dark matter halos are described with elliptical NFW profiles.

The critical curves for a source at redshift~4.8 from the two models are shown in Figure~\ref{fig:M0647_CC_LI}. Due to many bright foreground stars, we also see the corresponding bright blue-green diffraction spikes around them. The green-colored linear features (known as ``claws'' and ``Dragon's Breath Type II'') near lensed image~5.1, 12.3, and 21.2, are artifacts caused by the presence of bright stars far from the field of view~\citep[also see section 2.4 in][]{2022arXiv221014123H}. Fortunately, none of these artifacts lie near the giant arc, which is the focus of our current work.

\begin{figure}[t]
    \centering
    \includegraphics[scale=0.4]{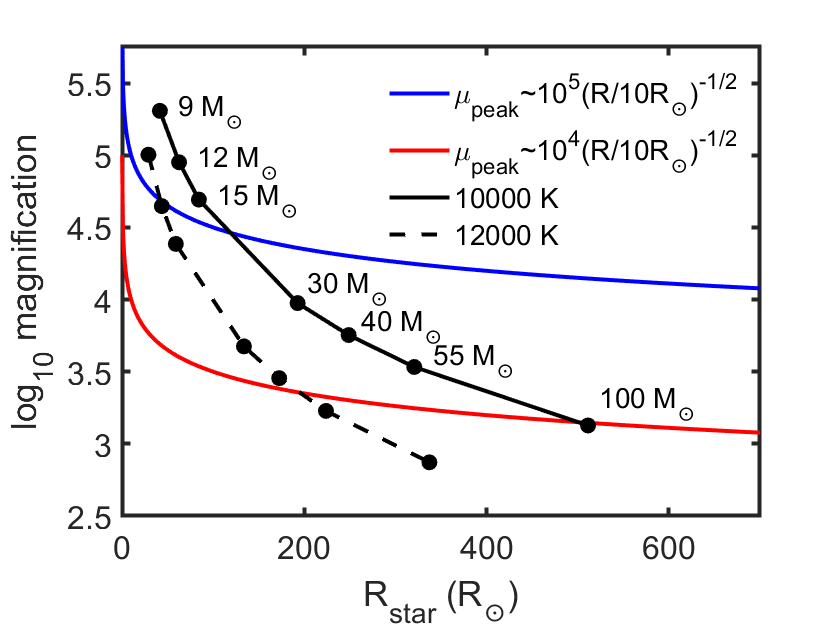}
    \caption{Radius-magnification limits for highly magnified stars. Red and blue lines represent the peak magnifications reached by micro-caustic crossings as a function of the star's radius \citep{2017ApJ...850...49V,Oguri2018}, for the two scenarios implied by the location of the \texttt{star-1} and \texttt{star-2}, respectively. Black filled circles along the solid and dashed black lines indicate the relation between stellar radius for $T_\mathrm{eff}\approx 10000$~K (\texttt{star-1}) and $T_\mathrm{eff}\approx 12000$~K (\texttt{star-2}) models of different initial mass from the BoOST \citep{Szecsi22} SMC-metallicity stellar evolutionary tracks, and the magnification required for these to match the apparent brightness of the two candidates. For \texttt{star-1}, only stellar models with initial masses $\gtrsim 20\ {\rm M}_\odot$ on the solid black line fall below the blue radius-magnification line, and are thus considered plausible, whereas stars smaller than this mass limit fall above the blue line, indicating they require too large magnifications for their predicted sizes. Similarly, for \texttt{star-2} only the stars with masses $\gtrsim50\ {\rm M}_\odot$ on the dashed black line fall below the red line and remain viable candidates. In the case of the stellar evolutionary tracks used here, stars with initial masses above 100 $M_\odot$ never reach temperatures as low as those inferred by the SED fits to our two candidates.} 
    \label{fig:mu_R_relation}
\end{figure}

\section{Highly Magnified Star(s) $\lowercase{\rm at}$ $\lowercase{z}\simeq4.8$}
\label{sec:LensedStars}

The \emph{JWST} imaging of the giant arc ($z\simeq4.8$) hosting the two-star candidates is marked by a white dashed box in Figure~\ref{fig:M0647_CC_LI}. The same giant arc in the six different NIRCam filters is shown in Figure~\ref{fig:z5_stamp}. In each image, we show the position of the lensed star candidates, \texttt{star-1} and \texttt{star-2}, by the green and red arrows, respectively. A color image of the same arc is shown in Figure~\ref{fig:z5_color}. In the color image, the position of \texttt{star-1} and \texttt{star-2}, (ra, dec) = ($6^{\rm hr}48^{\rm m}00.3732^{\rm s}$, $+70^\circ 14' 57\farcs948$) and ($6^{\rm hr}48^{\rm m}00.3320^{\rm s}$, $+70^\circ14'57\farcs761$), respectively, are shown by green and white stars.  The photometry of \texttt{star-1} and \texttt{star-2} are shown in Table~\ref{tab:photometry}.

\subsection{Macro-magnification}
To determine the position of the macro-critical curve independently of the lens models, we use a pair of the strongly lensed, multiply imaged clumps in the arc which are situated at the edges of the white lines drawn along the arc in Figure~\ref{fig:z5_color}. Using the ratio of distances between the two clumps on each side of the critical curve~(i.e., distance between 2.1a and 2.1b divided by distance between 2.2a and 2.2b), and the distance between the counter images of the innermost clump~(i.e., distance between 2.1a and 2.2a), the position of the macro-critical curve can be estimated and is shown by the yellow dashed line drawn perpendicular to the arc\footnote{One can also use the flux ratios of counter images of strongly lensed clumps to estimate the critical curve position independently.}. The critical curves predicted that way, although only a first-order approximation, are consistent with the lens models, offering greater confidence that these objects lie very near to the critical curves. According to this model-independent estimation of the critical curve, both \texttt{star-1} and \texttt{star-2} are situated on the saddle side, at a distance of~$<0\farcs1$ and~$\sim0\farcs32$ from the macro-critical curve. According to our \texttt{dPIEeNFW} model, \texttt{star-1} sits ``on'' (i.e., within $\sim$half the spatial resolution limit from) the macro-critical curve, whereas \texttt{star-2} lies at a distance of~$\sim0\farcs18$ from the macro-critical curve. The \texttt{Glafic} model gives a distance of~$\sim0\farcs25$ and~$\sim0\farcs50$ for \texttt{star-1} and \texttt{star-2} from the macro-critical curve, respectively. The macro-magnification~($|\mu|$) value at the position of \texttt{star-1}~(\texttt{star-2}) is~$\sim600$~($\sim90$) and~$\sim190$~($\sim50$) according to \texttt{dPIEeNFW} and \texttt{Glafic} lens models, respectively. 

\subsection{Photometry}

\input{star_photometry.tex}

To measure the photometry of the lensed star candidates, we follow a similar procedure as that described in~\citet{Welch2022EarendelJWST}. For each one of the filters, pixels associated with the star are identified using the clumps segmentation map provided by \texttt{NoiseChisel} and \texttt{Segment}~\citep{akhlaghi2015}. It is important to note that we run \texttt{Segment}, disabling any kernel convolution, to better identify the pixels that correspond to the star, thus avoiding any artificial extension of the region due to smoothing. We call these pixels "the star region". Then, we interpolate everything that surrounds the lensed star (the arc, the light from nearby galaxies, the intracluster light, and the sky) iteratively, using the interpolation algorithm included in the pipeline \texttt{CICLE}~\citep{jimenez-teja2018}. This algorithm places apertures in the star region and its surroundings randomly. If an aperture is not fully contained in the star region, its pixels are substituted by the median of the values outside the star region. Repeating this process iteratively, we cover the entire star region with apertures, interpolating all the pixels outside-in. The flux of the star is measured by subtracting the interpolated image from the original one, in a circular aperture of radius~$0\farcs3$. This approach has the advantage of minimizing the impact of a potential mis-estimation of the sky in the final photometry since this is the difference of two images this term cancels out. Because the apertures of the interpolation are placed randomly, we can get different values of the flux of the star in different realizations. We thus run the interpolation algorithm 100 times for each filter, and the standard deviation of the 100 measurements is included in the error budget along with the photometric error. Finally, the fluxes are corrected for encircled energy.

\subsection{Comparison with stellar models}
The photometry of \texttt{star-1} and \texttt{star-2} is shown by black solid-points in the middle and bottom panels of Figure~\ref{fig:star1}. The top panel shows the relevant filter response curves. For each candidate, we present stellar spectral energy distributions~(SEDs) based on the \citet{Lejeune97} set of stellar atmosphere spectra for three different effective temperatures ($T_\mathrm{eff}$) which all provide acceptable fits to the data given the error bars (blue, green and red lines, with the green line indicating the best fit). While the details of stellar atmosphere spectra depend on additional parameters such as metallicity and surface gravity, the coarse sampling of our photometric data points in practice only allow us to constrain $T_\mathrm{eff}$ (determined by the relative shape of the SED) and $\mu L$ (which determines the absolute scaling required to match the observed fluxes). Throughout this fitting exercise, we treat these as independent parameters. In the SED fits plotted, we have adopted stellar atmosphere spectra with metallicity [M/H] $=-1$ and surface gravity in the $\log(g)=2$--2.5 range, under the assumption that the redshift is $z=4.8$ and that the observed SED is unaffected by dust reddening. From the plot, we can see that a star with a temperature of $T_\mathrm{eff} \sim9,000 - 10,000$~K, i.e., in the transition from A- to B-type stars, provides a good fit to the observed photometry of \texttt{star-1} whereas a star with~$\sim12000$~K provides a reasonable fit for \texttt{star-2}, albeit which a more uncertain $T_\mathrm{eff}$ due to the significantly larger photometric errors. While the fits are relatively insensitive to the assumed metallicity, corrections for reddening due to circumstellar dust, or dust within the host galaxy of the star would shift both the inferred $T_\mathrm{eff}$ and luminosity to higher values.
When the $\mu L$ estimates from Figure~\ref{fig:star1} are combined with the macro-magnifications derived from the \texttt{dPIEeNFW} and \texttt{Glafic} lens models, it appears that $\log(L/L_\odot$) could be as high as 6.7-7.5 and 7.0-7.6 for the two stars respectively, suggesting that they may be evolved, extremely massive stars with initial masses in the 300-600 $M_\odot$ range \citep[e.g.][]{Szecsi22}. However, the magnification of each star could be significantly enhanced by microlensing by stars in the lensing cluster, which would significantly reduce the required intrinsic luminosities and stellar masses of these objects. Indeed, at temperatures of $T_\mathrm{eff}\sim 10000$ K, stars with $\log(L/L_\odot)> $ 5.5-6 would be in tension with the Humphreys-Davidson limit \citep{HumphreysDavidson79}, an empirical luminosity limit above which almost no $T_\mathrm{eff}\lesssim 15000$ K are known in the local Universe. 

To set a rough lower limit on the initial masses of stars that potentially could attain sufficient microlensing magnifications to match the observed fluxes of  \texttt{star-1} and \texttt{star-2}, we need to consider the likely sizes of these stars. Supergiants in the  $T_\mathrm{eff}\sim 10000$ K range can easily reach radii of several hundred $R_\odot$, and this limits the peak microlensing magnification that one can expect.
To estimate the peak magnification for \texttt{star-1} we assume it lies essentially on the critical curve (i.e. within the corrugated micro-caustic network), as suggested by our \texttt{dPIEeNFW} lens model.  We then use equation~(27) from~\citet{2017ApJ...850...49V}, which specifies the peak magnification in the corrugated network. Using the relevant lensing parameters such as a convergence value ($\kappa=0.55$), and the modulus of the gradient of $\kappa+\gamma$ \citep{2017ApJ...850...49V}, we obtain that the typical \emph{peak} magnification at the position of \texttt{star-1} is~$\sim[10^5, 10^4]\times(R/10~{R_\odot})^{-1/2}$ for a source with radius~$R$. Here we assumed that microlenses at the candidate position can contribute in the range~$[0.1\%, 1\%]$ of the total surface density~($\Sigma_{\rm tot}$) with a microlens mass of~$1~{\rm M_\odot}$. In Figure ~\ref{fig:mu_R_relation}, we use the peak magnification-star radius relation to demonstrate that stellar evolutionary tracks that predict high-mass stars reaching temperatures of $T_\mathrm{eff}\approx 10000$~K during late stages of evolution suggest that a subset of such stars ($\sim 20$--100 $M_\odot$ for the models plotted) are sufficiently compact to be extremely magnified and observed by us. 
As mentioned in the introduction, so far, nearly all galaxy clusters observed with JWST led to the detection of lensed stars. In addition, as shown in~\citet{Welch2022EarendelHST} for `Earendel', the probability of detecting massive lensed stars~($M\sim100~{\rm M}_\odot$) is roughly 1 in every 25-50 lensed arcs depending on various parameters, As our lensed star candidate lies at somewhat smaller redshift and has an allowed mass range~[$\sim 20$--100 $M_\odot$] implying that above number can be used as the lower limit. Apart from that, we note that the microlens density values similar to $0.1\times\Sigma_{\rm tot}$ might be unlikely to occur. But we should keep in mind that we are only estimating the average peak magnification, and we can still get higher (or lower) peak values than the average value. Hence, we may get peaks equivalent to the average peak value corresponding to $0.1\times\Sigma_{\rm tot}$ even for somewhat larger values of microlens density.

As for \texttt{star-2}, we find it most likely lies farther away from the macro-caustic and outside the corrugated network, whose expected size is~$\lesssim100$~milliarcseconds. At the position of \texttt{star-2}, the typical peak magnification is thus estimated using equation~(26) from~\citet{Oguri2018}, which describes the peak magnification outside the corrugated network, i.e., in the low optical depth regime -- where it is assumed that the peak is due to a single microlens caustic crossing. With this, we find~$\sim10^4(R/10~{R_\odot})^{-1/2}$, for which we can use the red line shown in Figure~\ref{fig:mu_R_relation} to estimate what type of stars can get bright enough to be observed. We find that the preferable mass for \texttt{star-2} is~$\gtrsim50~{\rm M_\odot}$.

\subsection{What if not lensed stars?}
One interesting possibility we should also consider is that the candidates are lensed star clusters instead of individual stars. Since the stars are unresolved, i.e., we adopt the PSF FHWM of $0\farcs03$ as their maximum size, and assuming a tangential magnification of~$500$ taken from \texttt{dPIEeNFW} model at the position of \texttt{star-1}, the upper limit on the rest-frame size of the source is~$<0.38$~pc, making a star cluster less likely. 
For \texttt{star-2}, which lies farther from the critical curve, we would expect a similarly magnified counter-image, was it a star cluster. 
As shown in Figure~\ref{fig:z5_color}, we do observe a possible counter-image of \texttt{star-2} on the other side of the macro-critical curve. However, this possible counter-image is noticeably fainter compared to the \texttt{star-2}, and so it is unclear if it is indeed a counter-image. If this is a counter-image, and \texttt{star-2} is a star cluster, the flux anomaly might be explained, for example, in one of the following ways: (1) The saddle side image which we dubbed \texttt{star-2}, gets an additional magnification boost due to the presence of a subhalo near its position, making it brighter. (2) One or more of the stars in the star cluster on the saddle-side image is going through a microlensing event, making the saddle-side image brighter than the minima-side image. Future observations will help determine the nature of these sources. We refer the reader also to \citep{Welch2022EarendelHST, Welch2022EarendelJWST}, for additional information on lensing of star clusters versus stars.

The original selection for the star candidates here was based on their compact size, proximity to the critical curves, symmetry arguments (in particular -- lack of counter images), and supported further by the SED fit. Nevertheless, other possible explanations should be acknowledged. We discuss here whether these may be small persistent objects at the redshift of the cluster, or at other low-redshift; or some transient phenomena, in the cluster, or at the source. The break near~$\sim2.1\mu{\rm m}$ in the measured SED of the candidates, seen in Figure~\ref{fig:star1}, matches very well the rest-frame Balmer break of A/B-stars at redshift~4.8 (rest frame wavelength of~$\sim0.3646\mu{\rm m}$). Owing to this break, the possibility that the candidates are interlopers at lower redshift or in the galaxy cluster -- such as compact galaxies, star clusters, or even transient phenomena such as supernovae  -- seems unlikely, because we do not expect a break at $\sim2.1\mu{\rm m}$ for typical objects at the cluster's redshift. 
We note that (as mentioned above) in Figure~\ref{fig:star1}, the presence of Balmer break for \texttt{star-1} is convincing whereas for \texttt{star-2} it is not due to the large error-bars.

Transient phenomena in the source galaxy, which in principle should be pondered as well given the lack of counter images, also seem unlikely: The expected observed time-delay between the observed candidates and their expected counter-images on the outer side of the macro-critical curve is~$<+0.42$~days where the `+' sign indicates that the observed candidates are trailing images so that images outside the macro-critical curve should have appeared up to a few hours before. Hence, only optical transients that last less than a few hours are possible candidates. If the candidates were some type of stellar explosions such as (kilo)novae or supernovae, we should have also detected their counter-images. The non-detection of counter-images on the outer side of the macro-critical curve thus allows us to discard any transient lasting more than~$0.42$~days in the observe frame.
This also includes a stellar-mass black hole accreting mass from an asymptotic giant branch (AGB) companion \citep[see][]{Windhorst2018} as such objects are not expected to show the observed break, and their timescale should be longer.

\section{Conclusions}
\label{sec:conclusion}

In this work, we report two highly magnified lensed star candidates detected in the \emph{JWST}/NIRCam imaging of MACS0647 acquired through the \emph{JWST} cycle~1 GO program~(program ID: 1433; PI: Dan Coe). These candidates were observed in a giant arc at a redshift of~$z_{phot}\simeq4.8$, making them the second farthest lensed star candidates after Earndel~\citep[$z\simeq6.2$;][]{Welch2022EarendelHST, Welch2022EarendelJWST} known to date. From a combination of magnification constraints and SED fiting, the estimated temperatures for the two stars are $\sim10,000$~K and $\sim12,000$~K, respectively. Using stellar evolutionary tracks, we find that stars with masses~$\gtrsim20~{\rm M_\odot}$ and~$\gtrsim50~{\rm M_\odot}$ are viable candidates for \texttt{star-1} and \texttt{star-2}, respectively, assuming peak magnifications inferred from the analytical relations given in~\citet{2017ApJ...850...49V} and~\citet{Oguri2018}, appropriate for our cases.


Based on the SED fit, lensing arguments -- including magnification, or proximity to the critical curves, symmetry, and time-delay -- along with the absence of counter images, we suggest that \texttt{star-1} is very likely a lensed star. For \texttt{star-2}, we observe a possible, faint counter image on the minima-side of the macro-critical curve, which -- if true -- may suggest it is instead a star cluster. In such a case the flux ratio anomaly between the star cluster and its expected counter image would need to be explained, possibly by micro- or milli- lensing at its observed position. Some other possible objects are also considered and deemed here unlikely, although it should be acknowledged that there may be other fitting, known or unknown, types of interlopers not considered here. 

Assuming that the candidates are indeed lensed stars (or star complexes), we can expect future observations would show variations in their light curves on timescale of hours to days, depending on the size of the source and relative velocity between lens and source. The size of the fluctuations is determined mainly by the distance from the caustic, size of the star, and underlying macro model parameters. Planned spectroscopic observations in early 2023 will help us further deduce the nature of these candidates.

\acknowledgements

A.K.M., A.Z. and L.J.F. acknowledge support by grant 2020750 from the United States-Israel Binational Science Foundation (BSF) and grant 2109066 from the United States National Science Foundation (NSF), and by the Ministry of Science \& Technology, Israel. 
Y.J-T. acknowledges financial support from the European Union’s Horizon 2020 research and innovation programme under the Marie Skłodowska-Curie grant agreement No 898633, the MSCA IF Extensions Program of the Spanish National Research Council (CSIC), and the State Agency for Research of the Spanish MCIU through the Center of Excellence Severo Ochoa award to the Instituto de Astrofísica de Andalucía (SEV-2017-0709).
E.Z. acknowledge funding from the Swedish National Space Agency. 
J.M.D. acknowledges the support of projects PGC2018-101814-B-100 and MDM-2017-0765. 
B.W. acknowledges support from NASA under award number 80GSFC21M0002. 
A. A. acknowledges support from the Swedish Research Council (Vetenskapsr\aa{}det project grants 2021-05559). 
R.A.B gratefully acknowledges support from the European Space Agency (ESA) Research Fellowship. MB acknowledges support from the Slovenian national research agency ARRS through grant N1-0238. 
P.D. acknowledges support from the NWO grant 016.VIDI.189.162 (``ODIN") and from the European Commission's and University of Groningen's CO-FUND Rosalind Franklin program.
G.M. acknowledges funding from the European Union’s Horizon 2020 research and innovation programme under the Marie Skłodowska-Curie grant agreement No MARACHAS - DLV-896778.
R.A.W. acknowledges support from NASA JWST Interdisciplinary Scientist grants NAG5-12460, NNX14AN10G and 80NSSC18K0200 from GSFC.
\facilities{\textit{JWST}(NIRCam), \textit{HST}(ACS, WFC3)}
\software{
\texttt{Python}~(\url{https://www.python.org}), 
\texttt{NumPy}~\citep{harris2020array}, 
\texttt{AstroPy}~\citep{astropy:2018}, 
\texttt{Matplotlib}~\citep{Hunter:2007},
\texttt{GRIZLI}~\citep{Grizli},
\texttt{EAZY}~\citep{2008ApJ...686.1503B},
\texttt{Glafic}~\citep{2010PASJ...62.1017O},
\texttt{dPIEeNFW}~\citep{2015ApJ...801...44Z}
}

\input{lensed_systems.tex}

\bibliographystyle{aasjournal} 
\bibliography{MyBiblio}

\end{document}

%% file: star_photometry.tex
\begin{deluxetable}{cccc}
\label{tab:photometry}
\tablecaption{Photometry of the \texttt{star-1} and \texttt{star-2}.}
\tablewidth{\columnwidth}
\tablehead{
\colhead{Filter} & \colhead{\texttt{star-1}} & \colhead{\texttt{star-2}} \\
\colhead{(1)}    &     \colhead{(2)}         &       \colhead{(3)}}
\startdata
\hline
F115W & $29.351\pm0.604$ & $29.786\pm0.823$ \\
F150W & $29.028\pm0.600$ & $29.921\pm0.721$ \\
F200W & $28.311\pm0.241$ & $28.746\pm0.301$ \\
F277W & $27.343\pm0.199$ & $28.330\pm0.378$ \\
F356W & $27.795\pm0.244$ & $29.059\pm0.471$ \\
F444W & $27.996\pm0.327$ & $29.406\pm0.666$ \\
\vspace{-0.1in}
\enddata
\tablecomments{Column 1: \emph{JWST} filter name; Column 2 \& 3: Measured apparent magnitudes of \texttt{star-1} and \texttt{star-2} with $1\sigma$ error bars.}
\end{deluxetable}

%% file: lensed_systems.tex
\startlongtable
\begin{deluxetable*}{ccccc}
\label{tab:lensed}
\tablecaption{Multiple image systems in MACS0647.}
\tablewidth{\columnwidth}
\tablehead{
\colhead{ID} & \colhead{R.A.} & \colhead{Dec.} & \colhead{$z_{\rm phot}$}& \colhead{Comments} \\
\colhead{(1)} & \colhead{(2)} & \colhead{(3)} & \colhead{(4)} & \colhead{(5)}}
\startdata
    1.1    &  101.9660445 &  70.2558166 &  $2.13_{-0.18}^{+0.33}$    &   \citet{2011MNRAS.410.1939Z}   \\
    1.2    &  101.9522096 &  70.2399951 &  $1.85_{-0.02}^{+0.55}$    &   --                            \\
    1.3    &  101.9666816 &  70.2483074 &          --                &   --            \vspace{0.02in} \\
\hline
    2.1    &  102.0013086 &  70.2501887 &  $4.76_{-0.16}^{+0.07}$    &   \citet{2011MNRAS.410.1939Z}   \\
    2.2    &  102.0013546 &  70.2487492 &  $4.79_{-0.15}^{+0.07}$    &   --                            \\
    2.3    &  101.9941933 &  70.2393721 &  $4.72_{-0.16}^{+0.13}$    &   --                            \\
\hline
    3.1    &  101.9743730 &  70.2433988 &  $3.32_{-0.10}^{+0.08}$    &   \citet{2013ApJ...762...32C}   \\
    3.2    &  101.9724519 &  70.2426198 &  $2.78_{-0.04}^{+0.49}$    &   --                            \\
\hline
    4.1    &  101.9281515 &  70.2493187 &  $2.46_{-1.34}^{+0.04}$    &   \citet{2013ApJ...762...32C}   \\
    4.2    &  101.9289077 &  70.2456830 &  $1.86_{-0.03}^{+0.66}$    &   --                            \\
    4.3    &  101.9389774 &  70.2571705 &  $2.05_{-0.27}^{+0.41}$    &   --            \vspace{0.02in} \\
\hline
    5.1    &  101.9209434 &  70.2514986 &  $6.33_{-0.26}^{+0.32}$    &   \citet{2013ApJ...762...32C}   \\
    5.2    &  101.9215289 &  70.2429026 &  $6.91_{-0.53}^{+0.37}$    &   --            \vspace{0.02in} \\
\hline
     6.11  &  101.9821971 &  70.2432514 &  $10.62_{-0.21}^{+0.27}$   &   \citet{2013ApJ...762...32C}   \\
     6.12  &  101.9713212 &  70.2397022 &  $10.58_{-0.37}^{+0.32}$   &   --                            \\
     6.13  &  101.9810192 &  70.2605628 &  $10.22_{-0.62}^{+0.50}$   &   --                            \\
     6.21  &  101.9821500 &  70.2433072 &          --                &   \citet{2022arXiv221014123H}   \\
     6.22  &  101.9711412 &  70.2397047 &          --                &   --                            \\
     6.23  &  101.9808954 &  70.2605925 &          --                &   --                            \\
     6.31  &  101.9827102 &  70.2438447 &  $0.52_{-0.06}^{+9.67}$    &   \citet{2022arXiv221014123H}   \\
     6.32  &  101.9697561 &  70.2393690 &  $9.19_{-6.97}^{+1.12}$    &   --                            \\
     6.33  &  101.9803691 &  70.2604279 &  $10.62_{-0.86}^{+0.92}$   &   --            \vspace{0.02in} \\
\hline
    7.1    &  101.9621250 &  70.2555278 &  $2.04_{-0.19}^{+0.18}$    &   \citet{2013ApJ...762...32C}   \\
    7.2    &  101.9488750 &  70.2397778 &          --                &   --                            \\
    7.3    &  101.9528750 &  70.2499444 &  $2.15_{-0.09}^{+0.17}$    &   --            \vspace{0.02in} \\
\hline
    8.1    &  101.9525417 &  70.2543889 &  $2.45_{-0.29}^{+0.08}$    &   \citet{2013ApJ...762...32C}   \\
    8.2    &  101.9472500 &  70.2534722 &  $2.31_{-0.24}^{+0.12}$    &   --                            \\
    8.3?   &  101.9433492 &  70.2385144 &  $2.37_{-0.30}^{+0.23}$    &   --            \vspace{0.02in} \\
\hline
    9.1    &  101.9324583 &  70.2501111 &  $5.74_{-0.24}^{+0.27}$    &   \citet{2013ApJ...762...32C}   \\
    9.2    &  101.9374167 &  70.2397778 &  $5.77_{-0.16}^{+0.40}$    &   --                            \\
    9.3    &  101.9544167 &  70.2604722 &  $5.94_{-0.26}^{+0.29}$    &   --            \vspace{0.02in} \\
\hline
    10.1   &  101.9196004 &  70.2490478 &  $7.34_{-0.16}^{+0.36}$    &   \citet{2015ApJ...801...44Z}   \\
    10.2   &  101.9205483 &  70.2448550 &  $7.33_{-0.11}^{+0.31}$    &   --            \vspace{0.02in} \\
\hline
    11.1   &  101.9783943 &  70.2530223 &  $1.98_{-0.22}^{+0.69}$    &   \citet{2015ApJ...801...44Z}   \\
    11.2   &  101.9798736 &  70.2491041 &          --                &   --                            \\
    11.3   &  101.9657264 &  70.2402669 &          --                &   --            \vspace{0.02in} \\
\hline
    12.1   &  101.9650223 &  70.2468672 &  $2.22_{-0.20}^{+0.19}$    &   \citet{2015ApJ...801...44Z}   \\
    12.2   &  101.9559227 &  70.2427413 &  $2.45_{-0.20}^{+0.06}$    &   --                            \\
    12.3   &  101.9677194 &  70.2583876 &  $2.14_{-0.13}^{+0.30}$    &   --            \vspace{0.02in} \\    
\hline
    13.1   &  101.9904001 &  70.2476587 &  $3.53_{-0.07}^{+0.21}$    &   New system                    \\
    13.2   &  101.9879212 &  70.2532128 &  $3.50_{-0.04}^{+0.24}$    &   --                            \\
    13.3   &  101.9749885 &  70.2378465 &  $3.50_{-0.04}^{+0.25}$    &   --            \vspace{0.02in} \\
\hline
    14.1   &  102.0023538 &  70.2438789 &  $2.86_{-0.12}^{+0.36}$    &   New system                    \\
    14.2   &  102.0020778 &  70.2436712 &  $3.21_{-0.20}^{+0.09}$    &   --                            \\
    14.3   &  102.0014310 &  70.2429844 &  $3.24_{-3.03}^{+0.15}$    &   --            \vspace{0.02in} \\
\hline
    15.1   &  101.9993248 &  70.2424316 &  $11.40_{-0.25}^{+0.21}$   &   New system                    \\
    15.2   &  102.0023951 &  70.2471152 &  $0.39_{-0.09}^{+0.19}$    &   --                            \\
    15.3   &  102.0023503 &  70.2502045 &  $3.38_{-0.25}^{+0.33}$    &   --            \vspace{0.02in} \\
\hline
    16.1   &  101.9455802 &  70.2488149 &          --                &   New system                    \\
    16.2   &  101.9447169 &  70.2488604 &  $6.69_{-2.20}^{+0.05}$    &   --                            \\
    16.3?  &  101.9682583 &  70.2605087 &  $6.71_{-0.11}^{+0.12}$    &   --                            \\
    16.4?  &  101.9523528 &  70.2392696 &  $6.70_{-0.07}^{+0.08}$    &   --            \vspace{0.02in} \\
\hline
    17.1   &  101.9211918 &  70.2463921 &  $2.46_{-0.32}^{+0.18}$    &   New system                    \\
    17.2   &  101.9211231 &  70.2460865 &          --                &   --                            \\
    17.3   &  101.9214513 &  70.2458354 &          --                &   --            \vspace{0.02in} \\
\hline
    18.1   &  101.9611602 &  70.2555089 &  $2.14_{-0.03}^{+0.03}$    &   New system                    \\
    18.2   &  101.9481108 &  70.2398067 &          --                &   --                            \\
    18.3   &  101.9521371 &  70.2501341 &  $2.15_{-0.04}^{+0.02}$    &   --            \vspace{0.02in} \\
\hline
    19.1   &  101.9898364 &  70.2484166 &  $3.73_{-0.73}^{+0.53}$    &   New system                    \\
    19.2   &  101.9881425 &  70.2524040 &  $3.60_{-0.55}^{+0.45}$    &   --                            \\
    19.3   &  101.9741549 &  70.2374749 &  $3.54_{-0.47}^{+0.42}$    &   --            \vspace{0.02in} \\
\hline
    20.1   &  101.9906731 &  70.2472096 &  $3.76_{-2.78}^{+0.60}$    &   New system                    \\
    20.2   &  101.9880141 &  70.2538524 &  $3.39_{-0.12}^{+0.22}$    &   --                            \\
    20.3   &  101.9763250 &  70.2382859 &  $3.31_{-1.52}^{+0.42}$    &   --            \vspace{0.02in} \\
\hline
    21.1   &  101.9513568 &  70.2503220 &  $2.86_{-0.18}^{+0.22}$    &   New system                    \\
    21.2   &  101.9623645 &  70.2562833 &  $2.73_{-0.12}^{+0.32}$    &   --                            \\
    21.3   &  101.9473538 &  70.2383398 &  $2.86_{-0.18}^{+0.26}$    &   --            \vspace{0.02in} \\
\hline
    22.1   &  101.9507687 &  70.2504387 &  $3.02_{-0.35}^{+0.15}$    &   New system                    \\
    22.2   &  101.9620375 &  70.2563121 &          --                &   --                            \\
    22.3   &  101.9471011 &  70.2383789 &          --                &   --            \vspace{0.02in} \\
\hline
    23.1   &  101.9897367 &  70.2442761 &  $3.05_{-0.27}^{+0.66}$    &   New system                    \\
    23.2   &  101.9890000 &  70.2436439 &  $2.74_{-1.94}^{+0.42}$    &   --            \vspace{0.02in} \\
\vspace{-0.1in}
\enddata
\tablecomments{Column 1: Lens system ID; Column 2 \& 3: R. A. and decl.; Column 4: \texttt{EAZY}~\citep{2008ApJ...686.1503B} photometric redshift with 95\% confidence interval estimated using \emph{HST} and \emph{JWST} observations; Column 5: System reference.}
\end{deluxetable*}